\documentclass[aps,pra,showpacs,floatfix,superscriptaddress,twocolumn]{revtex4-1}
\usepackage{graphicx}
\usepackage{dcolumn}
\usepackage{amsmath}
\usepackage{amssymb}
\usepackage{dsfont}
\usepackage{bm}
\usepackage{epstopdf}

\def\vec#1{{\boldsymbol{#1}}}
\def\<{\langle}
\def\>{\rangle}

\begin{document}

\title{Asymptotic dynamics of coined quantum walks on percolation
graphs}

\author{B. Koll\'ar}
\affiliation{WIGNER RCP, SZFKI, Konkoly-Thege Mikl\'os \'ut 29-33, H-1121 Budapest, Hungary}

\author{T. Kiss}
\affiliation{WIGNER RCP, SZFKI, Konkoly-Thege Mikl\'os \'ut 29-33, H-1121 Budapest, Hungary}

\author{J. Novotn\'y}
\affiliation{Department of Physics, Faculty of Nuclear Sciences and Physical Engineering, Czech Technical University in Prague, B\v
rehov\'a 7, 115 19 Praha 1 - Star\'e M\v{e}sto, Czech Republic}

\author{I. Jex}
\affiliation{Department of Physics, Faculty of Nuclear Sciences and Physical Engineering, Czech Technical University in Prague, B\v
rehov\'a 7, 115 19 Praha 1 - Star\'e M\v{e}sto, Czech Republic}


\date{\today}

\begin{abstract}
Quantum walks obey unitary dynamics: they form closed quantum systems.
The system becomes open if  the walk suffers from imperfections
represented as missing links on the underlying basic graph structure,
described by dynamical percolation. Openness of the system's dynamics creates decoherence, leading to strong mixing.
We present a method to analytically solve the asymptotic dynamics of coined, percolated quantum walks for a general graph structure. For the case of a
circle and a linear graph we derive the explicit form of the asymptotic
states. We find that a rich variety of asymptotic evolutions occur: not only the fully mixed state, but other stationary states, stable periodic and quasi-periodic oscillations can emerge, depending on the coin operator, the initial state and the topology of the underlying graph.
\end{abstract}

\maketitle

%%%%%%%%%%%%%%%%%%%%%%%%%%%%%%%%%%%%%%
Quantum walks (QWs) are natural generalizations of random walks, obeying unitary quantum evolution \cite{Aharonov1993}. The
possible applications of quantum walks range from quantum
information to modeling  transport phenomena \cite{KonnoReview}.
Quantum walks \cite{Kempe2003}, similarly to classical random walks, can serve as a tool to design efficient algorithms \cite{Magniez2011}.
At the same time, their conceptually simple definition with only slight modifications makes them a versatile model for a variety of
physical phenomena, motivating intensive research to understand their properties in recent years. In a series of very recent experiments
\cite{Experiments}
several aspects of quantum walks have been demonstrated. The next
generation of quantum walk experiments aims at demonstrating
possible fields of application of walks in solid state physics \cite{Rudner2009} or
open quantum dynamics \cite{Schreiber2011}.

Randomly removing edges from a graph one arrives at the problem of percolation
\cite{Percolation}. Percolation is perhaps the simplest, but a very generic model for disordered media.  In static percolation, the set of randomly
missing edges is fixed throughout the propagation, whereas in
dynamic percolation some random noise can change the connections
among the vertices in each time-step \cite{DynamicalPercolation}.
In the context of quantum walks, randomly broken links are a natural source of noise: in the quantum information setting they would correspond to errors, whereas in transport models they are a consequence of random variations of the medium.

The effect of broken links in coined quantum walks was first studied in \cite{Romanelli2005} as a general source of decoherence.
The long time behavior of the diffusion coefficient for the percolated walk on the line was calculated in \cite{Annabestani2010},
and its time evolution was numerically investigated in \cite{Kendon2010}.
The asymptotic limit for the global chirality distribution of a quantum walk on the line with percolation restricted to the half-line
was considered in \cite{Romanelli2011}.
Static percolation for coined quantum walks was studied in the context of spreading
\cite{Kendon2010}  and search efficiency \cite{Lovett2011} and, furthermore, for continuous-time quantum walks in \cite{Mulken2011}.

Time evolution of discrete time quantum walks is
unitary by definition, it can be specified by the coin and the step operators.
The iteratively defined evolution is then realized by repeated
application of the coin and the step operator maintaining deterministic evolution of the density operator. Randomness can enter
the quantum walk either by applying measurement \cite{Alagic2005,Kendon2007,Goswami2010} or introducing an external random variable. External disturbance may affect only the coin or only the step operator. Fluctuations in the coin operator usually lead to decoherence and the transition from ballistic
to diffusive motion. While a lot of interest was paid to changes of the
coin \cite{Ahlbrecht2010} much less is known about varying the step operator.
Percolation can be viewed as a perturbation of the step operator.
Other types of perturbation also lead to interesting directions, e.g. the realization of L\'evy flights
in QWs where connections with a distant vertex can appear, fundamentally modifying spreading properties \cite{Lavicka2011}.

In this Letter we resolve the asymptotic limits for coined quantum
walks on finite graphs with dynamical percolation. We obtain explicit
analytic forms for the asymptotic evolution for walks on a finite
linear graph and prove that generally there exist initial states
from which the completely mixed state cannot be reached. We
construct the attractor space of the superoperator governing the
evolution of the density operator. Beside various steady states, we
find stable asymptotic periodic oscillations as well as
quasi-periodic behavior.

Let us consider $d$-regular simple graphs $G(V, E)$
with a  finite number $N \equiv |V|$ of vertices. We introduce dynamical bond
percolation on this family of graphs as follows: Every edge in this
graph has the same independent $p$ probability of being present, and
the complementary $1-p$ probability of being broken (missing). In
every step of the quantum walk we randomly choose a
configuration of the edges $\mathcal{K} \subseteq E$. On each of
these configurations we define a unitary quantum walk by
combining the step operator and the coin operator. We assume the
coin operator to be always the same. Hence the
definition of the step operator is alternated. At positions with missing edges
we invoke the concept of reflection.

The Hilbert-space of discrete time coined quantum walks (QWs) on
percolated graphs defined above is the tensor product of the
$\mathcal{H}_P$ position space and the $\mathcal{H}_C$ coin space.
$\mathcal{H}_P$ is a $N$-dimensional Hilbert-space spanned by
orthonormal basis vectors corresponding to the vertices of the
graph, and $\mathcal{H}_C$ is $d$-dimensional, spanned by coin basis
states where $d$ is the regularity of the graph. The coin state
indicates the direction of the nearest neighbor vertex the walker is
displaced during the next time evolution step. Let us use the
notation $| a \rangle  \otimes | b \rangle \equiv | a, b \rangle $,
where $|a \rangle \in \mathcal{H}_P$ and  $|b \rangle \in
\mathcal{H}_C$.

The unitary time evolution of the QW on a given percolated graph is defined as
\begin{equation}
U_{\mathcal{K}} = S_{\mathcal{K}} \cdot \left( I_P \otimes C \right)\,,
\label{qwp_unitaries}
\end{equation}
where
\begin{eqnarray}
S_{\mathcal{K}}  & = &  \sum_{a, d} \left( \sum_{(a, a \oplus d) \in \mathcal{K}} | a \oplus d, d \rangle \langle a, d |  \right.  \nonumber\\
&  &  + \left. \sum_{(a, a \oplus d) \not\in \mathcal{K}}  \left( I_P \otimes R \right) | a, d \rangle \langle a, d |
  \right) \,,
\end{eqnarray}
and $C$ is the so called coin operator from the $U(d)$ group.
By $a \oplus d$ we denote the next neighbor of vertex $a$ in the direction $d$, and $R$ (reflection) is in general a traceless permutation matrix ensuring the unitarity of $S_{\mathcal{K}}$. In the examples considered in the present paper, $R$ further simplifies to a simple transposition matrix.
The effect of $S_{\mathcal{K}}$ could be considered as conditional displacement if the actual edge is present
and reflection if the edge is broken.

The randomness is introduced in the process by choosing different
edge configurations $\mathcal{K}$ at each step, and thus random
application of different unitary step operators. We denote the
probability of choosing a configuration $\mathcal{K}$, and
corresponding unitary steps $U_{\mathcal{K}}$ by
$\pi_{\mathcal{K}}(p) $. Since this process involves classical
randomness, the state of the walker is best described by a density
operator $\hat{\rho}(n)$, after $n$ steps. The complete state of the
system changes during time evolution according to
\begin{equation}
\label{time:evo:rho}
\hat{\rho}(n+1)=\sum_{\mathcal{K}} \pi_{\mathcal{K}} (p) U_{\mathcal{K}}
\hat{\rho}(n)
U^{\dagger}_{\mathcal{K}} \equiv \Phi \left(\hat{\rho}(n)\right),
\end{equation}
where we introduced the linear superoperator $\Phi$.

Numerical calculation of the superoperator in Eq.
(\ref{time:evo:rho}) over all possible $ |\mathcal{K} | = \sqrt{2^{d
\cdot N}}$  edge configurations by brute force methods would have an
exponential computational cost as a function of the number of
vertices $N$. In the following we show that this cost can be
reduced, requiring resources scaling polynomially with $N$. From now
on we use the short notation for the matrix elements $X^{s, c}_{t,
d} = \langle s, c | X  | t, d \rangle$. Utilizing Eq.
(\ref{time:evo:rho}) for matrix elements, and changing the sum over
elements with the sum over configurations we arrive at the
expression
\begin{equation}
\hat{\rho}^{s, c}_{t, d} (t+1) =   \sum_{a,b,q,r}  \hat{\rho}^{a, b}_{q, r}(t) \left( \sum_{\mathcal{K} \subseteq E} \pi_{\mathcal{K}}(p)\,  {U_{\mathcal{K}}}^{s,c}_{a,b} \,{U_{\mathcal{K}}^{*}}^{t, d}_{q, r} \right) \,.
\end{equation}
The second sum is over all possible configurations. The number of
configurations $| \mathcal{K} |$ grows exponentially with $N$.
Studying matrix elements of $U_{\mathcal{K}} = S_{\mathcal{K}} \cdot
C$, we recognize that only the elements of neighboring vertices are
nonzero: $S_{\mathcal{K}}$ is sparse. Consequently, the seemingly expensive
second sum could be taken just on edges between neighboring sets of
vertices $\xi$, thus $\mathcal{K} \subseteq \xi \subseteq E$. 
Since every vertex has $d$ neighbors, all $S_{\mathcal{K}}$ operators map a single vertex to one of the $d$ neighboring vertices. Thus a single run of the second sum, restricted to the set $\xi$, contains only $2^{2d} = 4^{d}$ additions in the worst case.
The first
summation is $\mathcal{O}(N^2)$ (polynomially) dependent on the
number of vertices, thus computation of a single time evolution step
is reduced to the polynomial regime with respect to $N$. Moreover,
the superoperator $\Phi$ is also a sparse operator, and the cost of
its construction is reduced to the polynomial regime as well.

\begin{figure}[tb!]
\begin{tabular}{p{0.24\textwidth}p{0.24\textwidth}}
\includegraphics[width=0.23\textwidth]{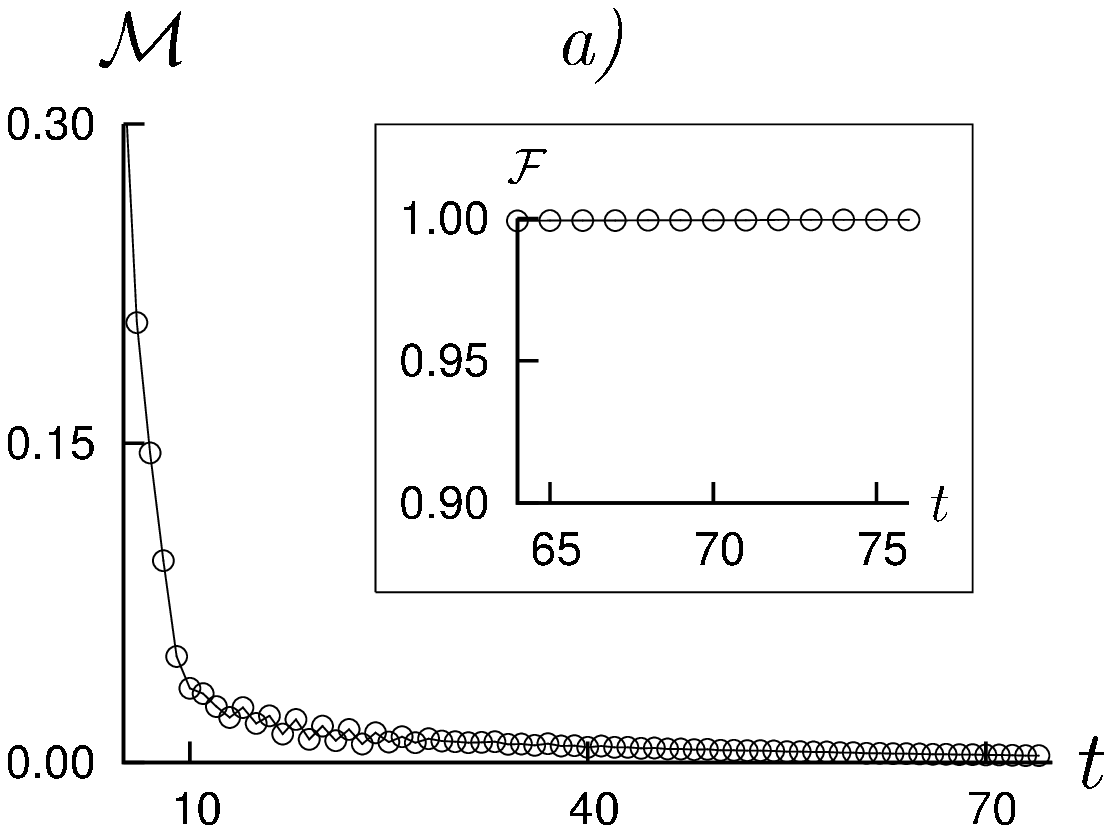} &
\includegraphics[width=0.23\textwidth]{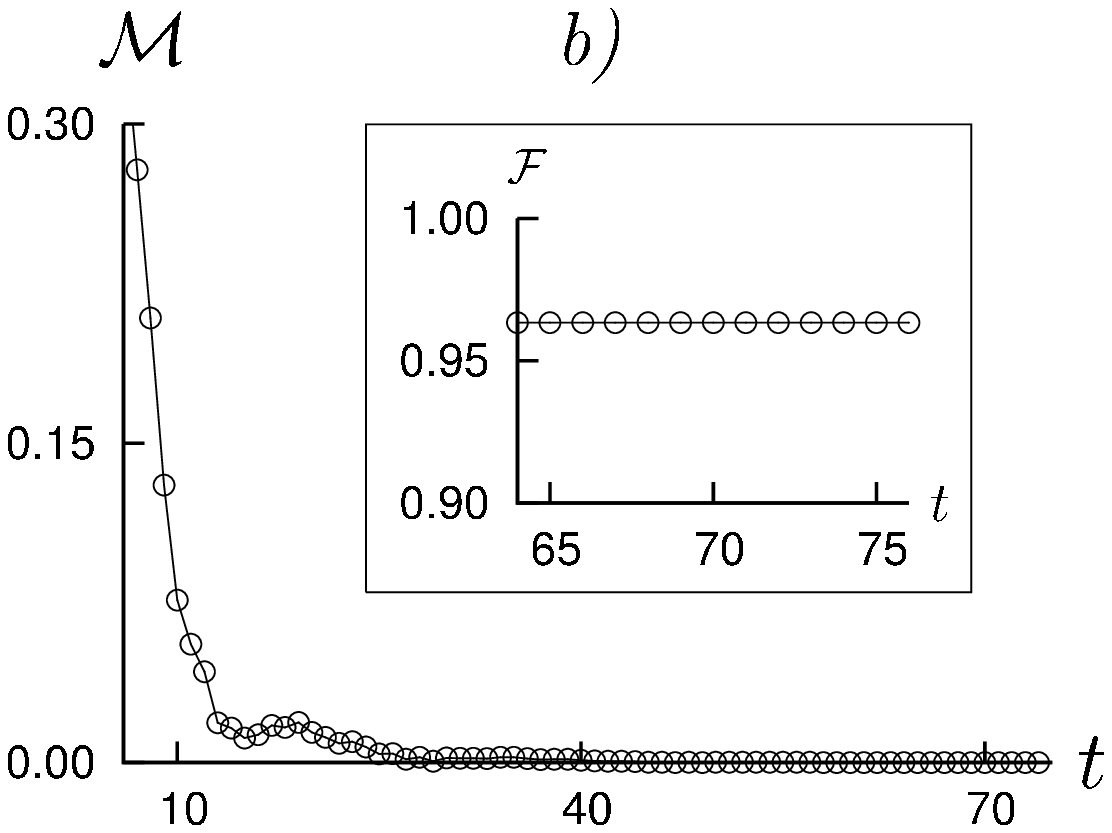} \\
\vspace{1pt}\includegraphics[width=0.23\textwidth]{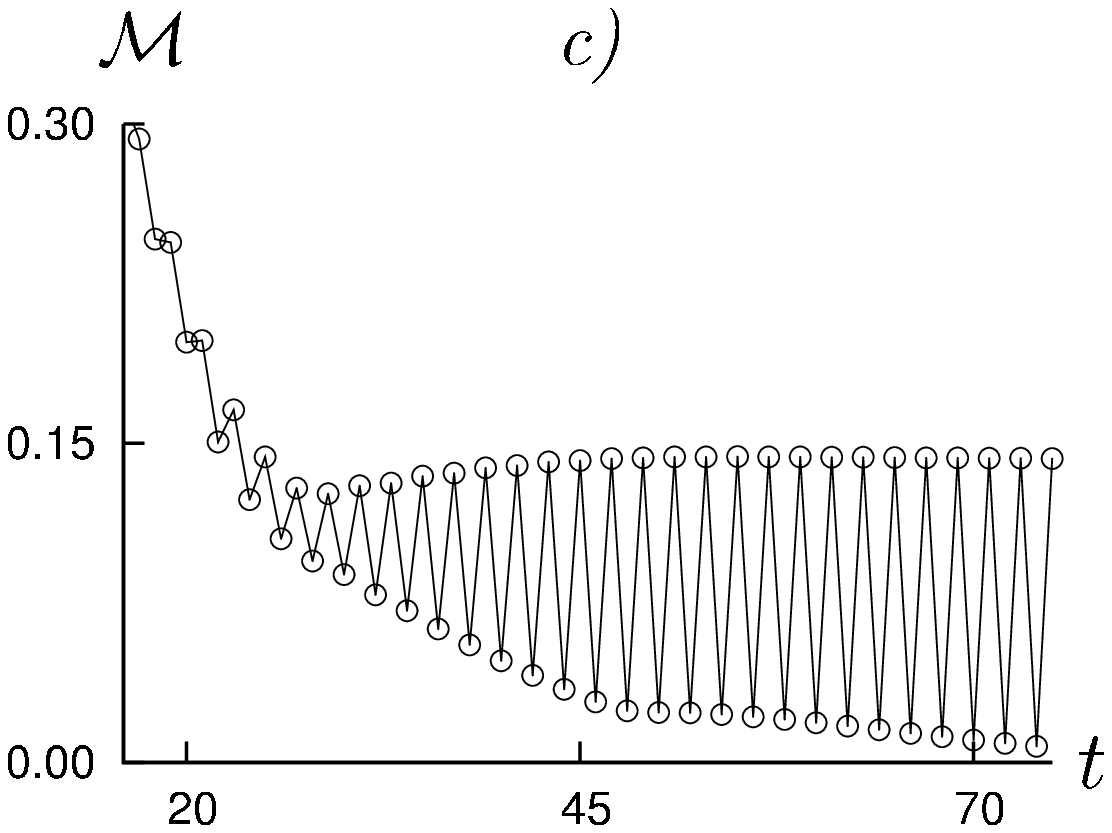} &
\vspace{1pt}\includegraphics[width=0.23\textwidth]{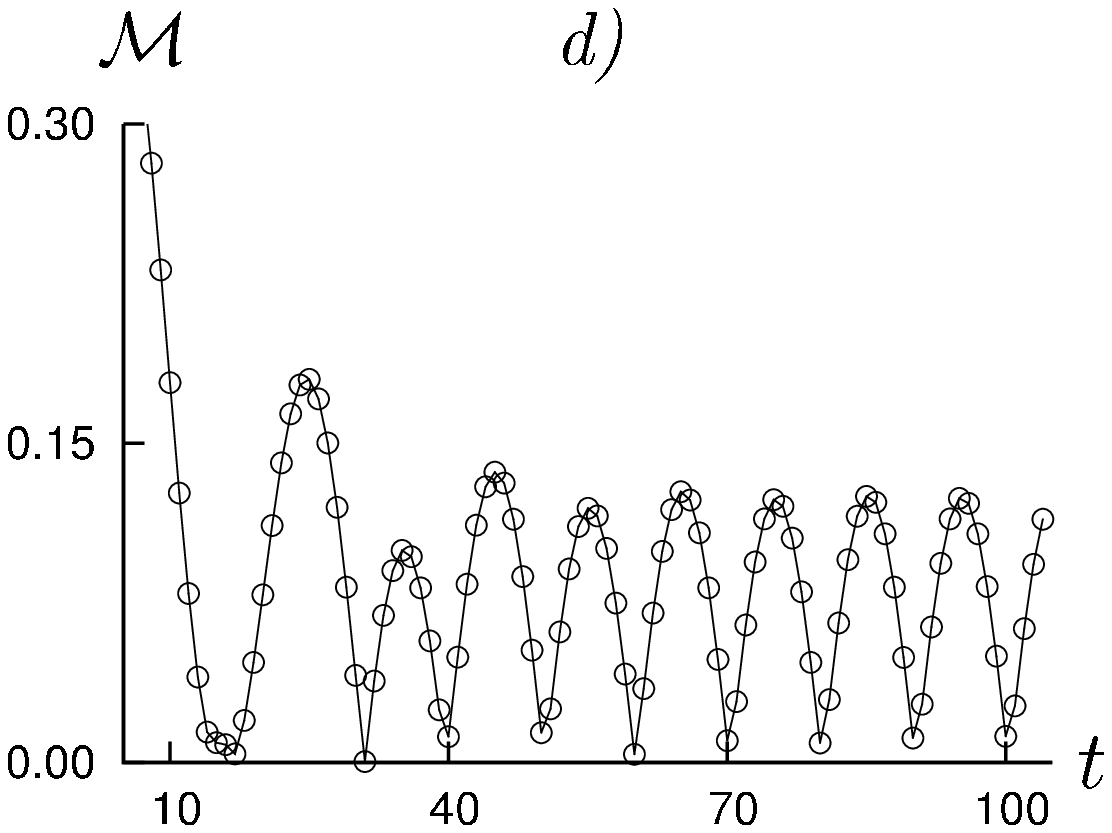}
\end{tabular}
\caption{Distance of the joint coin-position
distributions from the uniform distribution $\mathcal{M}(p,r) \equiv \sum | p_i - r_i |$ for percolated quantum
walks in dependence on the number of iterations. Initial states are
parametrized according to Eq. (\ref{loc:init:state}). {\it a)}
Hadamard walk on the 7-cycle with $\theta=\pi/2,\phi=-\pi/2$. Inset:
quantum fidelity with respect to the totally mixed state. {\it b)}
The same walk as in {\it a)} on the 8-cycle. The steady state here
is not totally mixed. {\it c)} 7-linear graph with
$\theta=-\pi/2,\phi=-\pi$. An asymptotic limit cycle occurs. {\it
d)} Special walk on the 8-cycle using the coin $C(\pi/2,\sqrt{2})$
(see Eq. (\ref{def:coin:1D})) with the same initial state as in {\it
c)} . The asymptotic behavior becomes quasi-periodic. In all cases
$p=0.5$, the points are connected for visual aid. } \label{Figure1}
\end{figure}

Decoherence caused by percolation leads to mixing of states, and one
might naively expect that asymptotically it reaches the completely
mixed state on a finite graph. Examining a particular system,
however, reveals that this is not the case. The joint position-coin
limit distribution is not necessarily uniform, one can obtain
(quasi) periodic behavior, as shown in Fig.~\ref{Figure1}.

%%Asymptotics%%%%%%%%
The asymptotic dynamics of the map $\Phi$ given by (\ref{time:evo:rho}) can be described 
via spectral decomposition \cite{Novotny2010}.
The process is somewhat similar to finding the eigenvalues and eigenvectors of a matrix,
but physically the resulting eigenoperators are not necessarily valid density operators.
First, we have to find all possible solutions of the equations
\begin{equation}
U_{\mathcal{K}} X_{\lambda,i} U_{\mathcal{K}}^{\dagger} = \lambda X_{\lambda,i}\,, ~{\rm with}~\, |\lambda|=1 
\,, ~~{\rm for~all}~\mathcal{K}\subseteq E\,,
\label{eigenmatrices}
\end{equation}
such that they form a maximal orthonormal set
 $\left\{ X_{\lambda,i}\right\}$.
The subspace spanned by $\left\{ X_{\lambda,i}\right\}$ is called the attractor space,
which includes density operators. The index $i$ distinguishes different mutually orthonormal solutions
with a certain eigenvalue $\lambda$. The density operators, contained in the attractor space,
are the possible asymptotic states of the system. The next task is to take into account the initial
condition, represented by the density operator $\hat{\rho}_{0}$ at time $t=0$.
In terms of the attractors, the asymptotic evolution ($n \gg 1$) is then given by
\begin{equation}
\hat{\rho}(n) = \Phi^n (\hat{\rho}_{0}) = \sum_{i, | \lambda | = 1} \lambda^n \mathrm{Tr}(X_{\lambda,i}^{\dagger} \hat{\rho}_{0}) X_{\lambda,i}\,.
\label{asymptotics}
\end{equation}
In case the attractor space contains more than one valid density matrices representing different physical states, then asymptotic cycles may occur. Depending on the size and structure of the attractor space,
the asymptotic evolution either converges to a single state or leads
to a periodic or a quasi-periodic limit cycle. 
As it was proven in  \cite{Novotny2010}, the generalized eigenvectors corresponding to eigenvalues $\lambda$ with
$|\lambda | = 1$ are all in fact eigenvectors (the corresponding Jordan blocks are one-dimensional). Moreover, the phases of these eigenvalues  determine the eventual (quasi) periodicity of the
corresponding asymptotic cycle. Examining Eq. (\ref{eigenmatrices}), one can see that it contains an equation for every unitary operator $U_{\mathcal{K}}$. These operators, in turn, are determined by the possible graphs with nonzero probability. Except the extremal cases of  $p=0$ or $p=1$, all graph configurations occur with finite probability, thus the set of unitaries is always the same for $p$ greater than $0$ and smaller than $1$, therefore  the asymptotic dynamics is independent of $p$ for $0<p<1$.

Employing (\ref{qwp_unitaries}) we rewrite the set of eigenvalue equations (\ref{eigenmatrices}) into
\begin{equation}
\left( I_{P} \otimes C \right) X_{\lambda} ( I_{P} \otimes C^{\dagger}) = \lambda S_{\mathcal{K}}^{\dagger} X_{\lambda} S_{\mathcal{K}}\,,
\label{cond:coin}
\end{equation}
which allows us to find the attractors in two steps. Eq. (\ref{cond:coin}) represents a set of equations where the left hand side is always the same. First we have to
find a subspace of matrices solving conditions imposed by different
shift operators
$
S_{\mathcal{K}}^{\dagger} X_{\lambda} S_{\mathcal{K}} = S_{\mathcal{K}'}^{\dagger} X_{\lambda} S_{\mathcal{K}'}
$
for all pairs of different edge configurations $\mathcal{K},
\mathcal{K}'$. Once we restrict the set of attractors to the latter
subspace, transitivity implies that it is sufficient to resolve an equation of type
(\ref{cond:coin}) only for a single chosen edge configuration. The most
favorable choice is the graph where all edges are missing, resulting
in the condition
\begin{equation}
\left( I_{P} \otimes R C \right) X_{\lambda} ( I_{P} \otimes C^{\dagger} R^{\dagger}) = \lambda X_{\lambda}\,.
\label{coin_reflection_equation}
\end{equation}
Using a block form $X_{\lambda}=\sum_{s,t} |s\>\<t| \otimes X^{(s,t)}$ we rewrite (\ref{coin_reflection_equation}) into the set of $N^2$ block equations
of the same form
$ \left(RC\right) X^{(s,t)} \left( RC \right)^{\dagger} = \lambda X^{(s,t)}$,
which can be turned into an eigenvalue problem
\begin{equation}
(RC) \otimes (RC)^* \vec x^{(s,t)} = \lambda \vec x^{(s,t)}\,.
\label{coin_vec_equation}
\end{equation}
The vector $\vec x^{(s,t)}$ is defined via the relation $\<a,b|\vec
x^{(s,t)}\> \equiv \<a|X^{(s,t)}|b\>$. The importance of equation
(\ref{coin_vec_equation}) is twofold. First, it shows that the
spectrum of the operator $RC$ determines all possible candidates of
the attractor spectrum. Second, eigenvectors corresponding to a
given eigenvalue $\lambda$ of the operator $(RC) \otimes (RC)^*$
reveal conditions imposed by the coin operator $C$ on each block
matrix $X^{(s,t)}$ of a possible attractor $X_{\lambda}$.

The just described method in principle allows for solving the
percolation problem for an arbitrary graph. The form of the solution
indicates the possibility to have nonstationary asymptotics (the
eigenvalues can have arbitrary phases while being of unit
magnitude). This is a particularly interesting result in view of the
general tendency of open systems to have a unique, single asymptotic
state. To illustrate the power of our method we give the analysis of
the simplest nontrivial 1 dimensional case.

%%Examples: cycle and linear graph%%%%%%%%

We apply the proposed method to solve the dynamical percolation problem on two topologically
different finite graphs: the linear graph and the circle, both consisting of $N$ vertices.
The linear graph can be viewed as a circle with a single constantly broken link thus
realizing reflecting boundary conditions, or vice versa the circle can be viewed as a line with periodic boundary conditions.
The step operator takes the form
\begin{eqnarray}
S_{\mathcal{K}} & = & \sum_a \sum_{d = 0, 1}  \left( \sum_{(a, a \ominus (-1)^{d}) \in \mathcal{K}} | a \ominus (-1)^{d}, d \rangle \langle a, d |  \right.  \nonumber\\
 && \left. + \sum_{(a, a \ominus (-1)^{d}) \not\in \mathcal{K}}  |a, 1 - d \rangle \langle a, d| \right)\,.
\label{def:shift:1D}
\end{eqnarray}
Here the operations $\oplus$ and $\ominus$ denote modulo-$N$
operations in case of the cycle and ordinary operations plus and
minus in case of the linear graph. This reflects different boundary
conditions which take place in definitions of both graphs. The
reflection operator is chosen $R=\sigma_{x}$.

Let us consider a two parameter class of coins
\begin{equation}
C(\alpha,\beta) = \left(\begin{array}{r r} ie^{-i \alpha}
\sin{\beta} &  \cos{\beta} \\ \cos{\beta} & i e^{i \alpha}
\sin{\beta} \end{array} \right), \label{def:coin:1D}
\end{equation}
with $\beta \neq \pi/2$. The case $\beta= \pi/2$ leeds to degeneracy \cite{Kollar}.
This class of coin operators includes the Hadamard coin for $\alpha=\pi/2, \beta=\pi/4$.
Examining the structure of Eq. (\ref{coin_vec_equation}), one can first determine the spectrum of the matrix $\sigma_x C(\alpha,\beta)$, which has two different eigenvalues $e^{\pm i
\beta}$. Consequently, the possible solutions of Eq. (\ref{coin_vec_equation}), i.e. attractor spectrum, can only be a subset of the set $\{1, e^{\pm 2i\beta}\}$.

Following the recipe outlined after Eq. (\ref{cond:coin}), after some tedious but straightforward calculation one can explicitly determine the basis of the attractor space.
For the linear graph (reflecting boundary conditions) the dimension of the attractor space corresponding to eigenvalue $\lambda=e^{2i\beta}$ is one, its base reads
\begin{equation}
X^{s,c}_{t,d}=\frac{1}{2N}\exp{(i\alpha\delta)}(-1)^{t+d}\,.
\label{attractor_i}
\end{equation}
For the  conjugate eigenvalue, $\lambda=e^{-2i\beta}$, the base can be chosen as $Y=X^{\dagger}$. The attractor space corresponding to the third eigenvalue, $\lambda=1$,  is three dimensional, its base $\{Z^1,Z^2,Z^3\}$ can be chosen in the form
\begin{eqnarray}
Z^1 &=& I_{2N}/\sqrt{2N}\,,
\label{atractor_1a}\\
\left(Z^2\right)^{s,c}_{t,d}&=&  \exp{[i\alpha(\delta-1)]}(1-(-1)^{\delta})/2 \sqrt{2}N\,,
 \label{attractor_1b}\\
\left(Z^3\right)^{s,c}_{t,d}&=&
\exp{(i\alpha\delta)} (1+(-1)^{\delta})/ 2 \sqrt{2N}\,,
\label{attractor_1c}
\end{eqnarray}
with $\delta=s-t-c+d$.

In the case of the cycle (periodic boundary conditions) we have additional constrains
and the dimension of the attractor space will depend on the length of the graph. If
$N=2k+1$ and $\alpha=l\pi/N$ $(l \in \mathds{N})$ the attractor
space has dimension equal to $2$, all attractors correspond to
eigenvalue $\lambda=1$ and its base can be chosen as
$\{Z^1,1/\sqrt{2}(Z^2+e^{i\alpha(N-1)}Z^3)\}$. When $N=2k$ and
$\alpha=2l\pi/N$, the structure of the attractor space is the same
as for the percolated linear graph, for $N=2$ and $\alpha \neq l\pi$
the attractor space is two-dimensional, all attractors correspond to
$\lambda=1$ and its orthonormal base can be chosen as $\{Z^1,Z^3\}$.
For all other cases on the cycle we found that there is only one
trivial solution of $Z^1$. These results clearly specify the topology induced difference between the walk on the percolated
linear graph and circle. This method becomes far more involved when
going to higher dimensions. However, also in this case the method
offers analytic results \cite{Kollar}.
\begin{figure}[tb!]
\begin{tabular}{cc}
\includegraphics[width=0.32\textwidth]{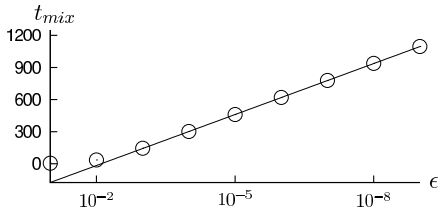} \\
\end{tabular}
\caption{ Position mixing time (circles) for a percolated
Hadamard walk on the 7-cycle as a function of distance threshold
$\epsilon$. Initially, the walker was localized on a vertex with
totally mixed coin state. The line represents our analytic
estimation Eq. (\ref{eq:t-mix-est}). } \label{Figure2}
\end{figure}
Let us illustrate the results obtained for coin $C(\alpha, \beta)$. We classify the initial states of the walk starting
from one given position $| x_{0} \rangle$, with respect to the asymptotic state it
reaches on a linear graph or on cycles with 5 dimensional attractor space. We use the standard
Bloch-sphere notation
\begin{equation}
\hat{\rho}_{0} = | x_{0}\rangle \langle x_0 | \otimes \left( P | \psi_{\theta,\phi} \rangle \langle \psi_{\theta,\phi} | +
\frac{1}{2} \left(1 - P \right) I_{C}  \right) \,, \label{loc:init:state}
\end{equation}
where $| \psi_{\theta,\phi} \rangle = \cos \frac{\theta}{2} | 0
\rangle + e^{i \phi} \sin \frac{\theta}{2} | 1 \rangle$. Examining
the attractor space, it is straightforward to see that the only
initial coin state leading to a single stationary asymptotic state,
which is not completely mixed, corresponds to $\theta=\pi/2$ and
$P>0$. The asymptotic steady state will be completely mixed only if
the initial coin state is totally mixed ($P=0$). The purity of the
asymptotic state ($ \mathrm{Tr}\left( \hat{\rho}_{as}^2 \right) =
\sum_{| \lambda | = 1} | \mathrm{Tr} \left\{ X_{\lambda}
\hat{\rho}_{0} \right\} |^2$) can be explicitly calculated for
localized initial states, yielding
\begin{equation}
\mathrm{Tr}\left( \hat{\rho}_{as}^2 \right) = \frac{1}{2 N} + \frac{P^2}{2 N ^2} \,.
\end{equation}
We immediately see that the most pure asymptotic state can be reached
from a pure ($P=1$) initial coin state.

The onset of the asymptotic regime can be characterized by the mixing time \cite{Kargin2010}.
One can employ the following definition for the position mixing time
$
t_{mix}(\epsilon) = \mathrm{inf}\left\{ t: \mathcal{M}\left[ p_P (t'), p_P(\infty) \right] \leq \epsilon , \, \forall \, t' \geq t  \right\}
$,
where $p_P(t)$ is the position probability distribution of a QW at time $t$, and $\mathcal{M}(p,r) \equiv \sum | p_i - r_i |$  is the Manhattan distance.
If one assumes that the slowest decaying states correspond to the largest  $|\lambda| \neq 1$ eigenvalue  of the superoperator $\Phi$ (in the simplest case there is only one such eigenvalue),
then one can estimate the mixing time
\begin{equation}
\label{eq:t-mix-est}
 t_{mix}^{(est)} (\epsilon) =  \log \epsilon / \log | \lambda' | - \log  \mathrm{|O|} /  \log | \lambda' |\,,
\end{equation}
where $\lambda'$ is the highest magnitude $|\lambda| \neq 1$
eigenvalue of superoperator $\Phi$. By $\mathrm{O}$ we denote the
overlap of the subspace corresponding to $\lambda'$ with the initial
state of the QW: $O = \mathrm{Tr}\left( X_{\lambda'}^{\dagger}
\hat{\rho}(0) \right)$. We illustrate this  result in
Fig.~\ref{Figure2}.

In summary, we presented an explicit method to solve the
asymptotic dynamics of percolated QWs.
Unlike in typical environment induced decoherence processes, in the considered case the asymptotic
open system dynamics can in general be oscillatory.
We emphasize that the asymptotic evolution does not depend on the percolation rate $p$ if $0<p<1$.
Already the analysis of the simple case of a linear-type graph provides nontrivial conclusions.
First, the possibilities for the asymptotic evolution on the linear graph are
fundamentally different from that of the cycle. This is a
consequence of the topology of the graph. One can expect the
occurrence of similar, but possibly more involved topological
effects for more complex graphs, e.g. for higher dimensional
lattices. Second, periodic boundary conditions lead to sensitivity
to the number of sites. 

In all the discussed cases, the position
distribution was asymptotically uniform, and asymptotic dynamics was
limited to the chirality degree of freedom. One can imagine that
more general graphs open the way for non-uniform asymptotic position
distributions.

We acknowledge support by MSM 6840770039, DI FNSPE CTU in Prague, the
Hungarian Scientic Research
Fund (OTKA) under Contract No. K83858 and
the Hungarian Academy of Sciences (Lend\"ulet Program,
LP2011-016).


\begin{thebibliography}{x}

\bibitem{Aharonov1993}
Y.~Aharonov et al.,
%\textit{ Quantum random walks},
Phys. Rev. A {\bf 48}, 1687 (1993).

\bibitem{KonnoReview}
N.~Konno, in
%Quantum walks,
\textit{Quantum Potential Theory, Lecture Notes in Mathematics},
U. Franz and M. Schurmann (Eds):
Vol. 1954, Springer, pp.309-452 (2008).

\bibitem{Kempe2003}
J. Kempe,
%\textit{Quantum random walks: An introductory overview},
Contemp. Phys., \textbf{44} (4), 307-327, (2003)

\bibitem{Magniez2011}
F. Magniez et al.,
%Ashwin Nayak, Jérémie Roland, and Miklos Santha
SIAM J. Comput. \textbf{40}, 142 (2011).

\bibitem{Experiments}
M.~Karski et al., Science \textbf{325}, 174 (2009); H.~Schmitz et
al., Phys. Rev. Lett. \textbf{103}, 090504 (2009); F.~Z\"ahringer et
al., Phys. Rev. Lett. \textbf{104}, 100503 (2010); A.~Schreiber et
al., Phys. Rev. Lett. \textbf{104}, 050502 (2010); M.~A. Broome et
al., Phys. Rev. Lett. \textbf{104}, 153602 (2010); A.~Peruzzo et
al., Science \textbf{329}, 1500 (2010).

\bibitem{Rudner2009}
M. S. Rudner, L. S. Levitov
%\textit{Topological Transition in a Non-Hermitian Quantum Walk}
Phys. Rev. Lett. \textbf{102}, 065703 (2009).

\bibitem{Schreiber2011}
A. Schreiber et al., Phys. Rev. Lett. \textbf{106}, 180403 (2011).

\bibitem{Percolation}
G. Grimmett, {\it Percolation}, Springer, Berlin (1999).

\bibitem{DynamicalPercolation}
J.~E.~Steif,
{\it A survey on dynamical percolation},
Fractal geometry and stochastics, IV, Birkhauser, 145-174. (2009).

\bibitem{Romanelli2005}
A.~Romanelli et al.,
%\textit{Decoherence in the quantum walk on the line},
Physica A \textbf{347}, 137 (2005).

\bibitem{Annabestani2010}
M. Annabestani et al,
%\textit{Decoherence in a one-dimensional quantum walk},
Phys. Rev. A \textbf{81}, 032321 (2010).

\bibitem{Kendon2010}
G. Leung et al.,
%\textit{Coined quantum
%walks on percolation graphs},
New J. Phys. \textbf{12}, 123018 (2010).

\bibitem{Romanelli2011}
A. Perez, A. Romanelli,
%\textit{Effects of broken links on the long-time behavior of quantum walks}
arXiv:1109.0122v1

\bibitem{Lovett2011}
N. B. Lovett et al.,
%\textit{The quantum walk search algorithm: Factors affecting efficiency},
arXiv:1110.4366v2

\bibitem{Mulken2011}
O. M\"ulken, A. Blumen,
%\textit{Continuous-time quantum walks: Models for coherent transport on complex networks}
Phys. Rep. \textbf{502}, 37 (2011).

\bibitem{Alagic2005}
G. Alagic, A. Russell,
%\textit{Decoherence in quantum walks on the hypercube},
Phys. Rev. A \textbf{72}, 062304 (2005)

\bibitem{Kendon2007}
V. Kendon,
%\textit{Decoherence in quantum walks � a review},
Math. Struct. in Comp. Sci. \textbf{17}, 1169-1220 (2007)

\bibitem{Goswami2010}
S. Goswami, P. Sen,
%\textit{Quantum persistence: A random-walk scenario},
Phys. Rev. \textbf{E} 81, 021121 (2010)

\bibitem{Ahlbrecht2010}
A. Ahlbrecht et al.,
%\textit{Asymptotic evolution of quantum walks with random coin},
arXiv:1009.2019v1

\bibitem{Lavicka2011}
H. Lavi\v cka et al.,
%\textit{Quantum Walk with Jumps},
Eur. Phys. J D {\bf 64}, 119
(2011)

\bibitem{Novotny2010}
J. Novotn\'y et al.,
%\textit{Asymptotic evolution of random unitary operators},
Cent. Eur. J. Phys. \textbf{8}, 1001 (2010); J. Phys. A: Math.
Theor. \textbf{42}, 282003 (2009).

\bibitem{Kollar}
B. Koll\'ar, T. Kiss, J. Novotny and I. Jex,
\textit{unpublished}

\bibitem{Kargin2010}
V. Kargin,
%\textit{Bounds for mixing time of quantum walks on finite graphs},
J. Phys. A: Math. Theor. \textbf{43}, 335302 (2010).

\end{thebibliography}
\end{document}